\documentclass[12pt]{article}

\usepackage{amsmath,amssymb}

\global\arraycolsep=1pt
\numberwithin{equation}{section}
\newcommand{\bel}[1]{\begin{equation}\label{#1}}                     
\newcommand{\bal}[1]{\begin{eqnarray}\label{#1}}                     
\newcommand{\be}{\begin{equation}}
\newcommand{\ee}{\end{equation}}

\newcommand{\im}{\mathrm{i}}
\newcommand{\ex}{\mathrm{e}}
\newcommand{\de}{\mathrm{d}}

\newcommand{\qq}{\qquad}

\renewcommand{\thefootnote}{\fnsymbol{footnote}}

\begin{document}

%
%
\begin{titlepage}

\begin{flushright}
       \normalsize
       OCU-PHYS 238  \\
       hep-th/0512124 \\
\end{flushright}

\vspace{5mm}

\begin{center}
{\Large \bf  Toric Sasaki-Einstein manifolds \\
and Heun equations}
\end{center}

\vspace{3mm}

\begin{center}
{Takeshi Oota$^1$\footnote{e-mail: \texttt{toota@sci.osaka-cu.ac.jp}}
and 
Yukinori Yasui$^2$\footnote{e-mail: \texttt{yasui@sci.osaka-cu.ac.jp}}}
\end{center}

\begin{center}
$^1${\sl Osaka City University, 
Advanced Mathematical Institute (OCAMI) \\
3-3-138 Sugimoto, Sumiyoshi, Osaka 558-8585, Japan}\\
\vspace{4mm}
$^2${\sl Department of Mathematics and Physics,
Graduate School of Science, \\
Osaka City University\\
3-3-138 Sugimoto, Sumiyoshi, Osaka 558-8585, Japan}\\
\end{center}


\begin{abstract}
Symplectic potentials are presented for a wide class of  
five dimensional toric Sasaki-Einstein manifolds, including
$L^{a,b,c}$ which was recently constructed by Cveti\v{c} et al. 
The spectrum of the scalar Laplacian on $L^{a,b,c}$ 
is also studied.
The eigenvalue problem leads to two Heun's differential
equations and the exponents at regular singularities
are directly related to the toric data.
By combining knowledge of the explicit symplectic
potential and the exponents, we show that
the ground states, or equivalently holomorphic functions,
have one-to-one correspondence 
with the integral lattice points in the convex polyhedral cone. 
The scaling dimensions of the holomorphic
functions are simply given by the scalar products of the Reeb vector
and the integral vectors, which are consistent with 
$R$-charges of the BPS states in the dual quiver gauge theories.

\end{abstract}

\vfill

\end{titlepage}

\renewcommand{\thefootnote}{\arabic{footnote}}
\setcounter{footnote}{0}


\section{Introduction}

Toric Sasaki-Einstein manifolds 
have been attracted much attention.
Five dimensional toric Sasaki-Einstein
manifolds $X_5$ can be used especially
as the type IIB backgrounds
whose near horizon geometries have the form $AdS_5 \times X_5$.
Recently, nontrivial infinite families of
toric Sasaki-Einstein manifolds were explicitly constructed
\cite{GMSW2,GMSW,CLPP,CLPP2}
and many new insights 
\cite{MS,BBC,BFHMS,HKW,MSY,MS2,BHOP,
GP,FHSU,BK,BBC2,BK05,FHMSVW,BFZ,BZ}
were obtained for the AdS/CFT correspondence
\cite{ADSCFT}.
CFT duals to the toric Sasaki-Einstein manifolds
are $\mathcal{N}=1$ superconformal quiver gauge theories
in $3+1$ dimensions \cite{QUIV,KW,MP}.
In these cases, both the geometry and the gauge theory
can be characterised by a common data, called a toric data.
On the CFT side, the formalism to determine the quiver gauge theory
from the toric data is now well established
and the extracted data can be beautifully encoded
in diagrams on a torus, called the brane tilings 
\cite{HK,FHKVW,FHMSVW,HV}.
However, on the gravity side, the procedure to obtain the metric
from the toric data is still not established.

A Sasaki-Einstein manifold
is a Riemannian manifold whose metric cone is
Ricci-flat K\"{a}hler i.e. 
Calabi-Yau. So far, 
many works on Calabi-Yau manifolds have been done
using the complex coordinates. However, Calabi-Yau or K\"{a}hler
manifolds are also symplectic. As is discussed in 
\cite{symp},
a real symplectic viewpoint can be useful in 
the K\"{a}hler toric geometry:
for example, explicit construction of extremal K\"{a}hler
metrics and spectral properties of the toric manifolds.
Therefore, some aspects of the toric Sasaki-Einstein manifolds
may become transparent in the symplectic coordinates \cite{MSY}.
The key object in the symplectic approach is the so called
symplectic potential, which is not well studied object
compared to the K\"{a}hler potential.
In the first part of this paper, we study the symplectic
potentials for various toric Sasaki-Einstein manifolds.

Then, we analyse the spectrum of the scalar Laplacian
for the toric Sasaki-Einstein manifolds 
$L^{a,b,c}$ \cite{CLPP,CLPP2} since 
the four dimensional base spaces of
$L^{a,b,c}$ 
are the most general \textit{orthotoric} K\"{a}hler-Einstein 
spaces
\cite{ACG,MS2}.
We will show that the eigenvalue problem 
leads to two Heun's differential equations \cite{ron},
and the exponents at the regular singularities 
are characterised by the toric data.
Although there is some progress in 
the construction of Heun functions 
(see, for example, \cite{val}), 
it is still difficult to solve
Heun's differential equations.
The determination of the spectrum of $L^{a,b,c}$ is also 
interesting as an eigenvalue problem of Heun's.
In this paper, we obtain some polynomial solutions,
which correspond to the ground states and the first excited
states.

Due to the AdS/CFT correspondence, there must be 
pairings between the ground states
and the BPS operators in the
dual quiver theories.
By combining knowledge of the symplectic potential
and the connection between the exponents and the toric data,
the corresponding BPS operators 
can be naturally identified.

This paper is organised as follows.
In the next section, after a brief review of the
symplectic approach to the toric Sasaki-Einstein manifolds,
we present the symplectic potentials for various cases.
In section $3$, we show how two Heun's differential
equations appear in the eigenvalue problem of the scalar Laplacians.
In section $4$, some polynomial type Heun functions are obtained.
Relations between the ground states,
holomorphic functions, and the BPS mesonic operators are argued
in section $5$.
Section $6$ is devoted to discussions.
Appendix treats some details on the symplectic potential for
$L^{a,b,c}$.

\section{Toric Sasaki-Einstein manifolds}

\subsection{Review of symplectic approach}

In this subsection, we briefly review the symplectic approach
to the toric Sasaki-Einstein manifolds according to \cite{MSY}.
Let $X_5$ be a five dimensional compact Riemannian manifold with 
a metric $g$.
The cone metric on $C(X_5) = X_5 \times \mathbb{R}_+$
is given by
$g_{C(X_5)} = \de r^2 + r^2 g$ ($r > 0$).
Sasaki-Einstein manifolds
can be defined as compact Einstein spaces whose metric cones 
are Ricci-flat K\"{a}hler (i.e. Calabi-Yau). 
The cone of a Sasaki-Einstein manifold
is also regarded as a symplectic manifold with the symplectic form
\be
\omega = \de \left( \frac{r^2}{2} \eta \right)
= r \de r \wedge \eta + \frac{r^2}{2} \de \eta.
\ee
Here $\eta$ is the contact one-form which is dual to the
distinguished Killing vector $B$, called the Reeb vector.
The symplectic coordinates $z = (\xi^1, \xi^2, \xi^3, 
\phi_1, \phi_2, \phi_3)$ on $C(X_5)$ can be chosen such that
\be
(r^2/2) \eta = \xi^i \de \phi_i =: \xi.
\ee
The one-form $\xi$ defined here is related to the symplectic form as
$\omega = \de \xi = \de \xi^i \wedge \de \phi_i$.

A five dimensional toric Sasaki-Einstein manifold is a Sasaki-Einstein
manifold with three $U(1)$ isometries : $G=T^3$. 
The isometry group $G$ acts
as translations for the coordinates $\phi_i$.
Let $\mathfrak{g}$ be its Lie algebra and $\mathfrak{g}^*$ 
the dual vector space of $\mathfrak{g}$.
The generators of $\mathfrak{g}$ are 
$\partial/\partial \phi_i$ $(i=1,2,3)$.
The moment map for the $T^3$-action is given by
\be
\begin{split}
\Phi: C(X_5) &\rightarrow \mathfrak{g}^* \cong \mathbb{R}^3, \\
z & \mapsto \xi = (\xi^1, \xi^2, \xi^3 ).
\end{split}
\ee
The image of the moment map is a convex polyhedral cone:
\be
\Phi \bigl( C(X_5) \bigr) 
= \{ \, \xi \in \mathfrak{g}^* \, |\, 
\langle v_A, \xi \rangle \geq 0, \ \ A=1,2,\dotsc, D \, \},
\ee
where
\be
v_A = \bigl( (v_A)_1, (v_A)_2, (v_A)_3 \bigr)
= ( v_A )_i \frac{\partial}{\partial \phi_i},
\qq
(v_A)_i \in \mathbb{Z}.
\ee
The pairing between the vector $v_A$ 
and the one form $\xi$ is given by
$\langle v_A, \xi \rangle = (v_A)_i \xi^i$.
Calabi-Yau conditions require that 
the vectors $v_A$ ($A=1,2,\dotsc, D$)
lie on a plane in the vector space $\mathfrak{g}$.
By appropriate choice of basis, the first component of these
vectors can be set to be one: $v_A = (1, w_A)$ with $w_A \in
\mathbb{Z}^2$. The set of vectors $\{ v_A \}$ is called a toric data.
Each subspace with $\langle v_A, \xi \rangle = 0$ ($A=1,2,\dotsc, D$)
is called a facet.
Note that the Sasaki-Einstein manifold is a $T^3$-fibration
over the hypersurface $\langle B, \xi \rangle = 1/2$
in the polyhedral cone.

In the symplectic coordinates, the cone metric
of the toric Sasaki-Einstein manifold can be expressed as
\be
g_{C(X_5)} = G_{ij}(\xi) \de \xi^i \de \xi^j
+ G^{ij}(\xi) \de \phi_i \de \phi_j,
\ee
\be
G_{ij} G^{jk} = \delta_i^k.
\ee
The metric is a cone if and only if $G_{ij}$ is homogeneous degree $-1$
in $\xi^k$.
The K\"{a}hler condition implies that $G_{ij}$
can be expressed by using a symplectic potential $G$:
\be
G_{ij}(\xi) = \frac{\partial^2 G(\xi)}{\partial \xi^i \partial \xi^j}.
\ee

The direct problem, to obtain the symplectic potential
and the toric data
from the metric, is not so formidable task.
On the other hand, the inverse problem, to obtain
the Sasaki-Einstein metric from the toric data, 
is very difficult.

Suppose we only know a toric data and do not know the corresponding 
Sasaki-Einstein
metric. In the symplectic approach, we should determine
the symplectic potential.
The moduli space of the symplectic potentials
for the toric Sasakian manifolds\footnote{
Sasakian manifolds can be defined as 
compact spaces whose metric cones are K\"{a}hler.}
can be written as $\mathcal{C}_0 \times \mathcal{H}(1)$.
Here $\mathcal{C}_0$ is the interior of the dual 
polyhedral cone
and it is identified with the space of the Reeb vectors 
$B=(B_1,B_2,B_3)$.  
Also, $\mathcal{H}(1)$ is the space of
smooth homogeneous degree one functions $h$ on the polyhedral cone.
The homogeneous one condition is required to guarantee that
$G_{ij}$ is homogeneous degree $-1$.

We must determine two quantities
$B$ and $h$ in order to obtain
the Sasaki-Einstein metric.
There is a well-established procedure to determine
the Reeb vector $B$ from the toric data.
It is the $Z$-minimisation. This procedure
is independent of the specification of $h$.
For any toric data, the first component of the Reeb vector is
fixed to be three: $B_1=3$.
  
The symplectic potential $G$ can be written as follows
\be
G = G^{\mathrm{can}} + G^B + h,
\ee
where
\be
G^{\mathrm{can}} = \sum_{A=1}^D \frac{1}{2} 
\langle v_A, \xi \rangle
\log \langle v_A, \xi \rangle,
\ee
\be
G^B = \frac{1}{2} \langle B, \xi \rangle
\log \langle B, \xi \rangle
- \frac{1}{2} \langle B^{\mathrm{can}}, \xi \rangle
\log \langle B^{\mathrm{can}}, \xi \rangle,
\ee
\be
B^{\mathrm{can}}:= \sum_{A=1}^D v_A.
\ee
The canonical part $G^{\mathrm{can}}$,
which is completely fixed by the toric data,
specifies the singular behaviour of $G$
at the facets.
By adding the second term $G^B$, the Reeb vector
shifts from $B^{\mathrm{can}}$ to $B$.
The third term $h \in \mathcal{H}(1)$ must be regular 
at the facets.

The Ricci-flatness condition is given by the Monge-Amp\`{e}re
equation
\be
\det \left( \frac{\partial^2 G}{\partial \xi^i \partial \xi^j}
\right) = \mathrm{const} \times
\exp\left( 2 \gamma^i \frac{\partial G}{\partial \xi^i} \right).
\ee
Here $\gamma^i$ are constants.
In order that the corresponding metric is smooth,
the constant vector $\gamma = ( \gamma^i)$ must be chosen as
\cite{symp}
\be
\gamma = (-1,0,0).
\ee

In general, $G^{\mathrm{can}}+ G^B$ 
is not a solution to the Monge-Amp\`{e}re equation.
So the discrepancy part $h$ is necessary.
In order to obtain the toric Sasaki-Einstein metric,
we should solve this quite non-linear partial differential
equation for $h$. 
In the subsequent subsections,
several solutions to the Monge-Amp\`{e}re equations
are presented.

\subsection{$C(T^{1,1})$ and $C(T^{1,1}/\mathbb{Z}_2)$}

The explicit homogeneous $T^{1,1}$ metric was constructed 
in \cite{CdlO}. The $T^{1,1}$ case is the first nontrivial example  
of toric Sasaki-Einstein/quiver duality \cite{KW}.

The toric data for $T^{1,1}$ is given by \cite{MP,MS}
\be
v_1 = (1,1,1), \qq
v_2 = (1,0,1), \qq
v_3 = (1,0,0), \qq
v_4 = (1,1,0).
\ee
The cone $C(T^{1,1})$ is the conifold.
By the $Z$-minimisation,
the Reeb vector is determined as $B=(3,3/2,3/2)$ \cite{MSY} and
$B^{\mathrm{can}} = \sum_{A=1}^4 v_A = (4,2,2)$.

The toric data for $T^{1,1}/\mathbb{Z}_2$ is given by \cite{MP}
\be
v_1 = (1, 0, 1), \qq
v_2 = (1, 1, 0), \qq
v_3 = (1, 2, 1), \qq
v_4 = (1, 1, 2).
\ee
The cone $C(T^{1,1}/\mathbb{Z}_2)$ in this case corresponds to the complex
cone over the zeroth Hirzebruch surface $\mathbb{F}_0$.
The Reeb vector is determined 
as $B = (3, 3, 3)$ and 
$B^{\mathrm{can}} = \sum_{A=1}^4 v_A = (4, 4, 4)$.

For $C(T^{1,1})$ or $C(T^{1,1}/\mathbb{Z}_2)$, 
the canonically constructed symplectic
potential $G = G^{\mathrm{can}} + G^B$ itself
is a solution to the Monge-Amp\`{e}re equation.
One can check that it indeed satisfies the Monge-Amp\`{e}re equation:
\be
\det \left( \frac{\partial^2 G}{\partial \xi^i \partial \xi^j}
\right) = \mathcal{C}
\frac{\langle B, \xi\rangle}
{\langle v_1, \xi \rangle \langle v_2, \xi \rangle 
\langle v_3, \xi \rangle  \langle v_4, \xi \rangle}
= \mathrm{const} \times
\exp\left( -2 \frac{\partial G}{\partial \xi^1} \right).
\ee
Here the constant $\mathcal{C}$ is $1/8$ for
$C(T^{1,1})$ and $1/2$ for $C(T^{1,1}/\mathbb{Z}_2)$.
The equation above is also valid for other choices of the toric data.

\subsection{Symplectic potentials of $C(Y^{p,q})$}

An infinite family of Sasaki-Einstein metrics 
with cohomogeneity one, called  
$Y^{p,q}$, was obtained in \cite{GMSW2}\footnote{
The positive integers $p$ and $q$ are assumed to be relatively prime.
This ensures that $Y^{p,q}$ is simply-connected, and thus 
diffeomorphic to $S^2 \times S^3$.}.

The toric data of $Y^{p,q}$ is given by \cite{MS}
\bel{YtdMS}
v_1 = (1, -1, -p), \qq
v_2 = (1, 0, 0), \qq
v_3 = ( 1, -1, 0 ), \qq
v_4 = (1, -2, -p+q).
\ee
From the $Z$-minimisation, the Reeb vector is determined as
\be
B = \left( 3 , -3, - \frac{3}{2} \left( p - q + \frac{1}{3 \ell}
\right) \right),
\ee
where
\be
\ell = \frac{q}{3q^2 - 2p^2 + p \sqrt{4p^2 - 3q^2}}.
\ee

We find that a homogeneous regular solution is given by
\be
h = \frac{1}{2} \langle B, \xi \rangle
\log \frac{\langle v_5, \xi \rangle}{\langle B, \xi \rangle}
+ \frac{1}{2} \sum_{A=1,3}
\langle v_A, \xi \rangle \log 
\frac{\langle B^{\mathrm{can}}, \xi \rangle}
{ \langle v_5, \xi \rangle }
+ \frac{1}{2} \sum_{A=2,4} 
\langle v_A, \xi \rangle \log
\frac{\langle B^{\mathrm{can}}, \xi \rangle}
{ | \langle v_6, \xi \rangle|},
\ee
where additional vectors $v_5$ and $v_6$
are defined by $v_5:= B - v_1 - v_3$,
$v_6:= - v_2 - v_4$,
and $B^{\mathrm{can}}= \sum_{A=1}^4 v_A$.
The symplectic potential of $C(Y^{p,q})$ is summed up into
a simple form
\bel{PotY}
\begin{split}
G(\xi) &= G^{\mathrm{can}} + G^B + h \\
&= \frac{1}{2} \sum_{I=1}^6 \langle v_{I}, \xi \rangle
\log | \langle v_I, \xi \rangle |.
\end{split}
\ee

One can check that the symplectic potential \eqref{PotY} indeed
satisfies the Monge-Amp\`{e}re equation:
\bel{MAY}
\det \left( \frac{\partial^2 G}{\partial \xi^i \partial \xi^j}
\right) = \frac{1}{32 a \ell^2}
\frac{\langle v_6, \xi \rangle^2}
{\langle v_1, \xi \rangle \langle v_2, \xi \rangle
\langle v_3, \xi \rangle \langle v_4, \xi \rangle
\langle v_5, \xi \rangle}
= \mathrm{const} \times
\exp\left( - 2 \frac{\partial G}{\partial \xi^1} \right),
\ee
where
\be
a = \frac{1}{2} - \frac{(p^2 - 3 q^2)}{4p^3}
\sqrt{4p^2 - 3q^2}.
\ee
This is one of our main results.
The expression of the symplectic potential \eqref{PotY}
and the Monge-Amp\`{e}re relation \eqref{MAY}
are valid not only for the choice \eqref{YtdMS} but also
for other choices of the toric data of $Y^{p,q}$ \cite{BFHMS,HKW}, 
which are related to \eqref{YtdMS}
by translations and $SL(2,\mathbb{Z})$ rotations.

\subsection{Symplectic potentials of $C(L^{a,b,c})$}

Cohomogeneity two generalisation of the $Y^{p,q}$ metric,
called the $L^{a,b,c}$ metric, was constructed in \cite{CLPP,CLPP2}.

The toric data of $L^{a,b,c}$ is given by \cite{FHMSVW}
\bel{toricDL}
v_1 = (1, 1, 0), \qq
v_2 = (1, ak, b), \qq
v_3 = (1, - al, c), \qq
v_4 = (1, 0, 0),
\ee
where the integers $k$, $l$ are chosen such that
$c k + b l = 1$. 
It holds that $av_1 - c v_2 + b v_3 - d v_4 = 0$
with $d=a+b-c$.

The Reeb vector $B$ can be obtained by using a solution
of certain quartic equation \cite{BK05}.
But, it is difficult to write
$B$ in a concise form. 
We do not write it
because the explicit form of the components $B_i$ 
is not relevant for solving the Monge-Amp\`{e}re equation.
For an implicit expression, see \eqref{ReebL}.
  
We solved the Monge-Amp\`{e}re equation for $L^{a,b,c}$ cases.
The solution is summarised as follows:
the symplectic potential for $C(L^{a,b,c})$ can be written as 
\bel{PotL}
\begin{split}
G(\xi) &= \frac{1}{2} \langle B, \xi \rangle
\log \langle B, \xi \rangle \\
& + \frac{1}{2} \sum_{m=1}^3 \langle v_{2m-1}, \xi \rangle
\log | x - x_m | 
+ \frac{1}{2} \sum_{m=1}^3 \langle v_{2m}, \xi \rangle
\log | y - y_m |,
\end{split}
\ee
where two auxiliary vectors in $\mathfrak{g}$ are 
introduced as
\be
v_5:= B - v_1 - v_3, \qq
v_6:= B - v_2 - v_4.
\ee
Here $x$ and $y$ are functions of $\xi^i$ and are defined implicitly by
\bel{vdefe}
\langle v_2, \xi \rangle = \frac{r^2}{4 \alpha}( \alpha - x) ( 1 - y),
\qq
\langle v_4, \xi \rangle = \frac{r^2}{4 \beta}(\beta - x) ( 1 + y),
\qq
\langle B, \xi \rangle = r^2/2.
\ee
Note that
\bel{vdefe2}
\langle v_6, \xi \rangle = \frac{r^2}{4 \alpha \beta}
x ( \alpha + \beta + (\alpha - \beta) y ).
\ee

The constants $\alpha$, $\beta$, $x_i$, $y_i$ ($i=1,2,3$)
obey certain consistency relations.
One can set the constants $y_i$ as follows:
\bel{yi}
y_1=1, \qq
y_2=-1, \qq
y_3 = \frac{\beta+\alpha}{\beta-\alpha}.
\ee
The following relations should be hold:
\bel{vdefo}
\begin{split}
\langle v_1, \xi \rangle &= \frac{r^2}{4}
\frac{(\alpha + \beta + (\alpha - \beta) y - 2 x_1)}
{(x_1-x_2)(x_1-x_3)}( x - x_1), \\
\langle v_3, \xi \rangle &= \frac{r^2}{4}
\frac{(\alpha + \beta + (\alpha - \beta) y - 2 x_2)}
{(x_2-x_1)(x_2-x_3)}( x - x_2), \\
\langle v_5, \xi \rangle &= \frac{r^2}{4}
\frac{(\alpha + \beta + (\alpha - \beta) y - 2 x_3)}
{(x_3-x_1)(x_3-x_2)}( x - x_3),
\end{split}
\ee
and further
\be
\de \langle B, \xi \rangle \wedge
\de \langle v_I, \xi \rangle \wedge 
\de \langle v_J, \xi \rangle 
= ( B, v_I, v_J) \de \xi^1 \wedge \de \xi^2 \wedge \de \xi^3
=: \frac{r^4 \rho^2}{16 \alpha \beta} K_{IJ} \de r^2 \wedge
\de x \wedge \de y,
\ee
\be
\de \langle v_{I}, \xi \rangle \wedge 
\de \langle v_{J}, \xi \rangle \wedge 
\de \langle v_{K}, \xi \rangle
= ( v_{I}, v_{J}, v_{K}) \de \xi^1 \wedge
\de \xi^2 \wedge \de \xi^3
=: \frac{r^4 \rho^2}{16 \alpha \beta} L_{IJK} \de r^2 \wedge
\de x \wedge \de y.
\ee
Here $\rho^2= (1/2)(\alpha + \beta) + (1/2)(\alpha - \beta) y - x$
and 
$(u,v,w)$ is the determinant of the $3 \times 3$
matrix whose rows are equal to the components of 
$u$, $v$ and $w$, respectively.
So we have a set of over-determined consistency conditions
for $\alpha$, $\beta$ and $x_i$. For example,
\bel{VL}
( v_I, v_J, v_K) L_{I'J'K'} = ( v_{I'}, v_{J'}, v_{K'} )
L_{IJK},
\qq I,J,K,I',J',K'=1,2,\dots, 6.
\ee
Some useful relations obtained from \eqref{VL} are
\bel{Lpara1}
\frac{x_1}{x_2} = - \frac{(v_1, v_5, v_6)}{(v_3, v_5, v_6 )},
\qq
\frac{x_3}{x_2} = - \frac{(v_1, v_5, v_6)}{(v_1, v_3, v_6)},
\ee
\bel{Lpara2}
\frac{\alpha - x_2}{x_2} = \frac{(v_2, v_3, v_4)}{(v_3, v_4, v_6)},
\qq
\frac{\beta - x_2}{x_2} = \frac{(v_2, v_3, v_4)}{(v_2, v_3, v_6)}.
\ee 
The constant $x_2$ can be scaled to be one. Then
these relations allow us to represent the set of parameters
in terms of the toric data.

The symplectic potential \eqref{PotL} 
satisfies the Monge-Amp\`{e}re equation and
reproduces the $L^{a,b,c}$ metric. 
See Appendix for details.
We note that in the $\alpha = \beta$ limit, 
the symplectic potential \eqref{PotL}
goes to the symplectic potential \eqref{PotY} for $Y^{p,q}$ 
up to irrelevant linear terms in $\xi^j$.

\subsection{The suspended pinch point}

As an application of the previous subsection,
we can explicitly write down the symplectic potential of 
the suspended pinch point (SPP) model. 
The toric data is 
given by \cite{MP}
\bel{TSPP}
v_0 = (1, 0, 0 ), \qq
v_1 = (1, -1, 0), \qq
v_2 = (1, 1, 0), \qq
v_3 = (1, 1, 1), \qq
v_4 = (1, 0, 1).
\ee
Note that 
$v_1 - v_2 + 2 v_3 - 2 v_4 = 0$.
The subset $\{v_1, v_2, v_3, v_4 \}$ is another
toric data for $L^{1,2,1}$. 
In this case, the blow-up vector 
$v_0$ is irrelevant for the symplectic potential and 
one can identify the metric for the SPP model with
the $L^{1,2,1}$ metric. 
Using the vectors $\{ v_A \}_{A=1}^4$ for the Z-minimisation,
the Reeb vector is determined as \cite{MSY}
\be
B = \left( 3, \frac{1}{2}( 3 - \sqrt{3}),
3 - \sqrt{3} \right).
\ee

Then, by setting $x_2=1$, \eqref{Lpara1} and \eqref{Lpara2}
fix the parameters as follows:
\be
x_1 = - \frac{1}{2} + \frac{1}{2} \sqrt{3}, \qq
x_2 = 1, \qq
x_3 = 1 + \sqrt{3}, \qq
\alpha = \frac{3}{2} + \frac{1}{2} \sqrt{3}, \qq
\beta = \sqrt{3}.
\ee
From \eqref{vdefe}, we have 
\be
\begin{split}
\frac{r^2}{2} &= \langle B, \xi \rangle 
= 3 \xi^1 + \frac{1}{2} ( 3 - \sqrt{3}) \xi^2
+ ( 3 - \sqrt{3}) \xi^3, \\
x &= \frac{(3/2)(1+\sqrt{3}) \xi^1 + (1/4)( -3 + 2\sqrt{3}) \xi^2
+ \sqrt{3} \xi^3 - (1/2)(3-\sqrt{3})\sqrt{D} }
{ 3 \xi^1 + (1/2) ( 3 - \sqrt{3}) \xi^2 + ( 3 - \sqrt{3}) \xi^3 }, \\
y &= \frac{
- ( 3 + 2 \sqrt{3}) \xi^1
- (2 + \sqrt{3}) \xi^2
- ( 1 + \sqrt{3} ) \xi^3
 + 2\sqrt{D}}
{ 3 \xi^1 + (1/2) ( 3 - \sqrt{3}) \xi^2 + ( 3 - \sqrt{3}) \xi^3 },
\end{split}
\ee
where
\be
D= 3( 2 + \sqrt{3} ) (\xi^1)^2 
+ \frac{1}{8}  (14 + 3 \sqrt{3}) (\xi^2)^2
+ 4 (\xi^3)^2 
+ \frac{3}{2}( 3 + \sqrt{3}) \xi^1 \xi^2
+ 3 ( 3 + \sqrt{3}) \xi^1 \xi^3
+ 4 \xi^2 \xi^3.
\ee
By substituting these expressions 
into \eqref{PotL}, we obtain the symplectic potential
for the SPP model.

\section{Scalar Laplacian and Heun's differential equations}

In this section, we study the scalar Laplacian for
the $L^{a,b,c}$ metric. 
The eigenvalue equation
can be separated into the ordinary
differential equations for each variables 
$x$, $y$, $\phi$, $\psi$, $\tau$.
The differential equations for the angle variables
$\phi$, $\psi$, $\tau$
can be solved in a trivial manner. For $x$ and $y$ variables,
these are Fuchsian type and are
shown to be Heun's differential equations \cite{ron}.

The $L^{a,b,c}$ metric is given by \cite{CLPP,CLPP2} (see also
Appendix)
\bel{Lmet}
\begin{split}
\de s^2 &= ( \de \tau + \sigma)^2 + \frac{\rho^2}{4 \Delta_x} \de x^2
+ \frac{\rho^2}{\Delta_{\theta}}
\de \theta^2 \\
& + \frac{\Delta_x}{\rho^2}
\left( \frac{\sin^2 \theta}{\alpha} \de \phi
+ \frac{\cos^2 \theta}{\beta} \de \psi \right)^2 
+ \frac{\Delta_{\theta} \sin^2 \theta \cos^2 \theta}
{\rho^2} \left( \frac{\alpha - x}{\alpha} \de \phi
- \frac{\beta - x}{\beta} \de \psi \right)^2,
\end{split}
\ee
where
\be
\sigma = \frac{(\alpha - x)}{\alpha} \sin^2 \theta \de \phi
+ \frac{(\beta - x)}{\beta} \cos^2 \theta \de \psi,
\ee
\be
\Delta_x = x ( \alpha - x)( \beta - x) - \mu
= (x-x_1)(x-x_2)(x-x_3),
\ee
\be
\rho^2 = \Delta_{\theta} - x,
\qq
\Delta_{\theta} = \alpha \cos^2 \theta + \beta \sin^2 \theta.
\ee
Three real roots of $\Delta_x$ are chosen such that
$0 < x_1 < x_2 < x_3$.
It is convenient to change the coordinate $\theta$ to
$y$ by setting $y=\cos 2\theta$.
The coordinates $x$ and $y$ have the ranges $x_1 \leq x \leq x_2$
and $-1 \leq y \leq 1$.  

The scalar Laplacian for the $L^{a,b,c}$ metric \eqref{Lmet}
is given by
\bel{sLap}
\begin{split}
\Box_{(5)}
&= \frac{4}{\rho^2} \frac{\partial}{\partial x} 
\left( \Delta_x \frac{\partial}{\partial x} \right) 
+ \frac{4}{\rho^2} \frac{\partial}{\partial y} 
\left( \Delta_y \frac{\partial}{\partial y} \right) 
+ \frac{\partial^2}{\partial \tau^2} \\
&+ \frac{\alpha^2 \beta^2}{\rho^2 \Delta_x}
\left(
\frac{(\beta-x)}{\beta} \frac{\partial}{\partial \phi}
+ \frac{(\alpha -x)}{\alpha} \frac{\partial}{\partial \psi}
- \frac{(\alpha -x)(\beta-x)}{\alpha \beta} 
\frac{\partial}{\partial\tau} \right)^2 \\
& + \frac{\alpha^2 \beta^2}{\rho^2 \Delta_y}
\left(\frac{(1+y)}{\beta} \frac{\partial}{\partial \phi}
- \frac{(1-y)}{\alpha} \frac{\partial}{\partial \psi}
- \frac{(\alpha - \beta)(1-y^2)}{2 \alpha \beta}
\frac{\partial}{\partial \tau} \right)^2.
\end{split}
\ee
Here $\Delta_y := (1-y^2) \Delta_{\theta} = (1/2)(1-y^2)
(\alpha+\beta +(\alpha-\beta)y)$.
By using \eqref{yi}, it also can be written as
\be
\begin{split}
\Box_{(5)}
&= \frac{\partial^2}{\partial \tau^2} \\
& +\frac{4}{\rho^2} \frac{\partial}{\partial x} 
\left( \Delta_x \frac{\partial}{\partial x} \right) 
+ \frac{\Delta_x}{\rho^2}
\left(
\frac{1}{x-x_1} v_1 + \frac{1}{x-x_2} v_3
+ \frac{1}{x-x_3} v_5
\right)^2 \\
&+ \frac{4}{\rho^2} \frac{\partial}{\partial y} 
\left( \Delta_y \frac{\partial}{\partial y} \right) 
+ \frac{\Delta_y}{\rho^2}
\left(
\frac{1}{y-y_1} v_2 + \frac{1}{y-y_2} v_4 
+ \frac{1}{y-y_3} v_6
\right)^2,
\end{split}
\ee
where
\be
v_1 = - \ell_1, \qq
v_3 = - \ell_2, \qq
v_5 = - \ell_3.
\ee
\be
v_2 = \frac{\partial}{\partial \phi}, \qq
v_4 = \frac{\partial}{\partial \psi}, \qq
v_6 = \frac{\partial}{\partial \tau}
- \frac{\partial}{\partial \phi}
- \frac{\partial}{\partial \psi}.
\ee
Here
\be
\ell_i = c_i \frac{\partial}{\partial \tau}
+ a_i \frac{\partial}{\partial \phi}
+ b_i \frac{\partial}{\partial \psi},
\ee
\bel{abc}
a_i = \frac{\alpha c_i}{x_i - \alpha}, \qq
b_i = \frac{\beta c_i}{x_i - \beta}, 
\qq
c_i = \frac{(\alpha - x_i)(\beta -x_i)}
{2(\alpha + \beta) x_i - \alpha \beta - 3 x_i^2}.
\ee
Note that
\be
a v_1 - c v_2 + b v_3 - d v_4 = 0, \qq
v_1 + v_3 + v_5 = B, \qq
v_2 + v_4 + v_6 = B.
\ee
Here
\be
B = \frac{\partial}{\partial \tau} = B_i \frac{\partial}{\partial \phi_i}
\ee
is the Reeb vector for $L^{a,b,c}$. More explicitly,
\bel{ReebL}
B = (B_1, B_2, B_3) 
= \left( 3, - \frac{1+ak a_1}{c_1}, - \frac{b a_1}{c_1} \right).
\ee

\subsection{Heun's differential equations}

The eigenfunctions of the scalar Laplacian,
$\Box_{(5)} \Psi = -E \Psi$,
have the form
\be
\Psi = \exp\left( \im N_{\tau} \tau
+ \im N_{\phi} \phi + \im N_{\psi} \psi \right)
F(x) G(y).
\ee
Here $N_{\tau}$, $N_{\phi}$, $N_{\psi}$ are constants.
They are related to constants $N^i$ as
$N_{\tau} \tau + N_{\phi} \phi + N_{\psi} \psi
= N^i \phi_i$. The coordinates $\phi_i$ are assumed to be
$2\pi$-periodic. Then $N^i \in \mathbb{Z}$ $(i=1,2,3$).
It is convenient to introduce 
the following $1$-form:
\bel{formN}
N:= N_{\phi} \de \phi + N_{\psi} \de \psi
+ N_{\tau} \de \tau
=N^i \de \phi_i.
\ee
For the toric data \eqref{toricDL}, the explicit relations
between these constants are given by
\be
N_{\phi} = \langle v_2, N \rangle = N^1 + ak N^2 + b N^3, \qq
N_{\psi} = \langle v_4, N \rangle = N^1, \qq
N_{\tau} = \langle B, N \rangle = B_i N^i.
\ee
$N_{\phi}$ and $N_{\psi}$ take values in integers,
while $N_{\tau}$ generally takes a value in $\mathbb{R}$.

The differential equations for $F$ and $G$ can be written as
\bel{HeunF}
\frac{\de^2 F}{\de x^2}
+ \left( \frac{1}{x-x_1} + \frac{1}{x-x_2}
+ \frac{1}{x-x_3} \right)
\frac{\de F}{\de x} + Q_x F = 0,
\ee
\bel{HeunG}
\frac{\de^2 G}{\de y^2}
+ \left( \frac{1}{y-y_1} + \frac{1}{y-y_2} + \frac{1}{y-y_3}
\right)
\frac{\de G}{\de y} + Q_y G = 0,
\ee
where
\be
Q_x = \frac{1}{\Delta_x}
\left( \mu_x -  \frac{1}{4} E x - \sum_{i=1}^3
\frac{\alpha_i^2 \Delta_x'(x_i)}{x-x_i} \right), \qq
Q_y = \frac{1}{ H_y}
\left( \mu_y - \frac{1}{4} E y
- \sum_{i=1}^3 \frac{\beta_i^2 H'(y_i)}{y-y_i}
\right).
\ee
Here
\bel{alphai}
\alpha_i := - \frac{1}{2}
( a_i N_{\phi} + b_i N_{\psi} + c_i N_{\tau}),
\ee
\bel{betai}
\beta_1 := \frac{1}{2} N_{\phi}, \qq
\beta_2 := \frac{1}{2} N_{\psi}, \qq
\beta_3 := \frac{1}{2} ( N_{\tau} - N_{\phi} - N_{\psi}),
\ee
\be
H_y:= \frac{2 \Delta_y}{\beta -\alpha}
= (y-y_1)(y-y_2)(y-y_3),
\ee
\be
\mu_x:= \frac{1}{4} C - \frac{1}{2} N_{\tau}
( \alpha N_{\phi} + \beta N_{\psi}) 
+ \frac{1}{4} (\alpha + \beta) N_{\tau}^2,
\ee
\be
\mu_y:= \frac{1}{2(\beta-\alpha)}
\left( -C + \left(\frac{\alpha + \beta}{2}\right) E + 2(\alpha N_{\phi}
+ \beta N_{\psi} )N_{\tau} 
-(\alpha+\beta) N_{\tau}^2 \right),
\ee
and $C$ is a constant.
These differential equations for $F$ and $G$ are
the Fuchsian type with four regular singularities
at $x = x_1, x_2, x_3, \infty$,
and at
$y = y_1, y_2, y_3, \infty$,
respectively.
Therefore, they are Heun's differential equations \cite{ron}.
For the first Heun's differential equation,
the exponents are
$\pm \alpha_i$ at $x=x_i$ ($i=1,2,3$), while $-\lambda$
and $\lambda+2$ at $x=\infty$.
For the second,
the exponents are
$\pm \beta_i$ at $y=y_i$ ($i=1,2,3$), while $-\lambda$
and $\lambda+2$ at $y=\infty$.
Here we put
$E = 4 \lambda ( \lambda + 2)$.

We find that the exponents at $x=x_i$ and $y=y_i$ 
are related to the toric data as
\bel{exx}
\langle v_1, N \rangle = 2\alpha_1, \qq
\langle v_3, N \rangle = 2\alpha_2, \qq
\langle v_5, N \rangle  = 2\alpha_3,
\ee
\bel{exy}
\langle v_2, N \rangle = 2\beta_1, \qq
\langle v_4, N \rangle = 2\beta_2, \qq
\langle v_6, N \rangle = 2\beta_3.
\ee 
These are important relations which will be useful to discuss 
properties of the eigenfunctions. 
Note that from \eqref{alphai} and \eqref{betai},
$\alpha_1 + \alpha_2 + \alpha_3 = \beta_1 + \beta_2 + \beta_3 =
(1/2) N_{\tau}$.

Heun's differential equations \eqref{HeunF} and \eqref{HeunG}
can be converted into the 
standard forms by the coordinate transformations
\be
\tilde{x} = \frac{x-x_1}{x_2-x_1}, \qq
\tilde{y} = \frac{y-y_1}{y_2-y_1},
\ee
with the rescalings\footnote{
Here,  for simplicity,  
we assume that $\alpha_1$, $\alpha_2$, $\beta_1$,
and $\beta_2$ are all positive.
All negative case can be
treated by replacing the powers in 
\eqref{rescale1} and \eqref{rescale2}:
$\alpha_i\rightarrow -\alpha_i$,
$\beta_i \rightarrow - \beta_i$. Otherwise various relations
between the constants $a_i$, $b_i$, and $c_i$ \eqref{abc}
cannot be used and the analysis
becomes very complicated. We will comment on this assumption later.
See section 5.} 
\bel{rescale1}
F(x) = \tilde{x}^{\alpha_1} ( 1- \tilde{x})^{\alpha_2}
(a_x - \tilde{x})^{\alpha_3} f(\tilde{x}), \qq
a_x:= \frac{x_3 - x_1}{x_2 - x_1},
\ee
\bel{rescale2}
G(y) = \tilde{y}^{\beta_1} ( 1 - \tilde{y})^{\beta_2}
( a_y - \tilde{y} )^{\beta_3} g(\tilde{y}),
\qq
a_y := \frac{y_3-y_1}{y_2-y_1}.
\ee
The resulting two standard form of Heun's differential equations 
are 
\be
\frac{\de^2 f}{\de \tilde{x}^2}
+ \left( \frac{\gamma_x}{\tilde{x}}
+ \frac{\delta_x}{\tilde{x}-1}
+ \frac{\epsilon_x}{\tilde{x}-a_x} \right)
\frac{\de f}{\de \tilde{x}}
+ \frac{\hat{\alpha} \hat{\beta} \tilde{x} - k_x }
{\tilde{x}(\tilde{x}-1)(\tilde{x}-a_x)}
f = 0,
\ee
\be
\frac{\de^2 g}{\de \tilde{y}^2}
+ \left( \frac{\gamma_y}{\tilde{y}}
+ \frac{\delta_y}{\tilde{y}-1}
+ \frac{\epsilon_y}{\tilde{y}-a_y} \right)
\frac{\de g}{\de \tilde{y}}
+ \frac{\hat{\alpha} \hat{\beta} \tilde{y} - k_y }
{\tilde{y}(\tilde{y}-1)(\tilde{y}-a_y)}
g = 0,
\ee
where
\be
\hat{\alpha}= - \lambda + \frac{1}{2} N_{\tau},
\qq
\hat{\beta}= 2 + \lambda + \frac{1}{2} N_{\tau}.
\ee
\be
\gamma_x= 2 \alpha_1 + 1, \qq
\delta_x= 2 \alpha_2 + 1, \qq
\epsilon_x= 2 \alpha_3 + 1,
\ee
\be
\gamma_y= 2 \beta_1 + 1, \qq
\delta_y= 2 \beta_2 + 1, \qq
\epsilon_y= 2 \beta_3 + 1,
\ee
and the accessory parameters are given by
\be
k_x = ( \alpha_1 + \alpha_3)( \alpha_1 + \alpha_3 +1)
- \alpha_2^2 
 + a_x \bigl[ ( \alpha_1 + \alpha_2)
( \alpha_1 + \alpha_2 +1) - \alpha_3^2 \bigr]
 - \tilde{\mu}_x,
\ee
\be
k_y = ( \beta_1 + \beta_3)( \beta_1 + \beta_3 +1)
- \beta_2^2 
+ a_y \bigl[ ( \beta_1 + \beta_2)
( \beta_1 + \beta_2 +1) - \beta_3^2 \bigr]
 - \tilde{\mu}_y.
\ee
Here
\be
\tilde{\mu}_x:= \frac{1}{x_2-x_1} \left( \mu_x
- \frac{1}{4} E x_1 \right), 
\qq
\tilde{\mu}_y:= \frac{1}{y_2-y_1} \left( \mu_y
- \frac{1}{4} E y_1 \right).
\ee
Note that the parameters $\hat{\alpha}$ and $\hat{\beta}$
are common for both Heun's differential equations.
With some work, the accessory parameters are summarised as follows:
\be
k_x =
\frac{1}{x_2-x_1}
(\tilde{C} -  \hat{\alpha} \hat{\beta} x_1 ) ,
\qq
k_y
= \frac{1}{\beta-\alpha}
(\tilde{C} - \hat{\alpha} \hat{\beta} \alpha ).
\ee
Here
\be
\tilde{C}:= \frac{1}{2} ( \alpha + \beta ) N_{\tau}
- \frac{1}{2} ( \alpha N_{\phi} + \beta N_{\psi} )
- \frac{1}{4} C.
\ee
In the following, $\tilde{C}$ will be treated as 
an undetermined constant. 

\subsection{Comments on the $Y^{p,q}$ limit}

For $\alpha = \beta$ limit, the $L^{a,b,c}$ metric reduces 
to the $Y^{p,q}$ metric. 
In this limit, $y_3$ goes to $\infty$ and 
the differential equation for the $y$-system \eqref{HeunG} becomes
\be
\frac{\de^2 G}{\de y^2}
+ \left( \frac{1}{y-1} + \frac{1}{y+1} \right)
\frac{\de G}{\de y} + 
\sum_{i=1}^2 \left( - \frac{\beta_i^2}{(y-y_i)^2}
+ \frac{Q^{(1)}_i}{y-y_i} \right) G = 0,
\ee
where
\be
Q_1^{(1)} = - Q_2^{(1)} = \frac{1}{8} ( \alpha^{-1} C - E
+ N_{\tau}^2 - 2 N_{\phi} N_{\psi} ).
\ee
This differential equation is the Fuchsian type 
with three regular singularities at $\pm 1, \infty$.
The exponents are $\pm \beta_i$ at $y=y_i$ ($i=1,2$)
and $-J$, $J+1$ at $y=\infty$.
Here $J$ is defined by
\be
J(J+1) = \beta_1^2 + \beta_2^2 - 2 Q_1^{(1)}
= \frac{1}{4} \left[
E - \alpha^{-1} C  - ( N_{\tau} - N_{\phi} - N_{\psi} )
(N_{\tau} + N_{\phi} + N_{\psi} ) 
\right].
\ee
One can solve the differential equation in terms
of Gauss' hypergeometric function:
\be
G(y) = ( y - y_1 )^{\beta_1} ( y - y_2 )^{\beta_2}
F\bigl(\tilde{\alpha}, \tilde{\beta}, \tilde{\gamma}; (1/2)(1-y)
\bigr),
\ee
where
\be
\tilde{\alpha} = \beta_1+ \beta_2 - J, \qq
\tilde{\beta} = \beta_1 + \beta_2 + J + 1, \qq
\tilde{\gamma} = 1 + 2 \beta_1.
\ee
When $\tilde{\alpha}$ is a non-positive integer,
the hypergeometric function becomes a Jacobi polynomial.
The constant $J$ here is the exact $SU(2)$-spin $J$ in \cite{KSY}.
 
\section{Polynomial solutions}

In this section, the spectrum of the scalar Laplacian \eqref{sLap} 
of $L^{a,b,c}$ is investigated.
The $L^{a,b,c}$ family contains $T^{1,1}$ and $Y^{p,q}$
as special cases. The case of $T^{1,1}$ was studied in \cite{gub} 
and the case of $Y^{p,q}$ in \cite{BHOP,KSY}.

The spectrum can be determined by solving the two Heun's
differential equations.
In principle, one can obtain
local power-series solutions of Heun's by solving
three-term recursion relations. 
But it is not easy to 
write the power series in a compact manner. 
Also, the local solutions should be regular functions 
at the regular singularities $0$ and $1$.  
Such regular solutions are called the Heun functions \cite{ron}.
The regular solutions of polynomial type appear when the
three-term recursion relations terminate at certain degree.
We study the polynomial solutions
for the two Heun's differential equations.

\subsection{Recursion relations}

For $\lambda = \frac{1}{2} N_{\tau} + n$ 
($n \in \mathbb{Z}_{\geq 0}$),
the parameters $\hat{\alpha}$ and $\hat{\beta}$ take values as
$\hat{\alpha} = - n$, 
$\hat{\beta} = N_{\tau} + n + 2$, and
a polynomial solution of degree $n$ may be possible:
\be
f(\tilde{x}) = \sum_{m=0}^{n} d_m \tilde{x}^m, \qq
g(\tilde{y}) = \sum_{m=0}^{n} e_m \tilde{y}^m.
\ee
Since the parameters $\hat{\alpha}$ and $\hat{\beta}$
are common for both Heun's differential equations,
the degree of both polynomials must be equal, if there exist.

The coefficients $d_m$ and $e_m$ satisfy the 
following three-term recursion relations:
\bel{Recx}
\begin{split}
- k_x d_0 + a_x \gamma_x d_1 &= 0, \\
P_m d_{m-1} - (Q_m^{(x)} + k_x) d_m + R_m^{(x)} d_{m+1} &= 0,
\qq (m=1,2,\dotsc, n-1), \\
P_n d_{n-1} - (Q_n^{(x)} + k_x) d_n &= 0.
\end{split}
\ee
\bel{Recy}
\begin{split}
- k_y e_0 + a_y \gamma_y e_1 &= 0, \\
P_m e_{m-1} - (Q_m^{(y)} + k_y) e_m + R_m^{(y)} e_{m+1} &= 0,
\qq (m=1,2,\dotsc, n-1), \\
P_n e_{n-1} - (Q_n^{(y)} + k_y) e_n &= 0.
\end{split}
\ee
Here
\be
\begin{split}
P_m &= ( m-1 + \hat{\alpha})( m - 1 + \hat{\beta}), \\
Q_m^{(x)} &= m \bigl( (m-1 + \gamma_x) (1 + a_x) + a_x
\delta_x + \epsilon_x \bigr), \\
Q_m^{(y)} &= m \bigl( (m-1 + \gamma_y) (1 + a_y) + a_y
\delta_y + \epsilon_y \bigr), \\
R_m^{(x)} &= ( m + 1) ( m + \gamma_x ) a_x, \\
R_m^{(y)} &= ( m + 1) ( m + \gamma_y ) a_y. \\
\end{split}
\ee
The requirement for existence of nontrivial solutions
can be written in the following form:
\bel{detM}
\det( M_x^{(n)} - k_x ) = 0, \qq
\det( M_y^{(n)} - k_y ) = 0.
\ee
Here $M_x^{(n)}$ and $M_y^{(n)}$ are $(n+1) \times (n+1)$
matrices whose matrix elements can be read 
from \eqref{Recx} and \eqref{Recy} respectively.
These relations \eqref{detM}
are two algebraic equations for one constant $\tilde{C}$
which is hidden in the accessory parameters $k_x$ and $k_y$.
They impose quite strong restrictions on $\tilde{C}$.

\subsection{Constant solutions ($n=0$)}

For $n=0$, let us set $\tilde{C}=0$. Then $k_x = k_y = 0$
and we have constant solutions with the eigenvalues
\bel{E0}
E = N_{\tau} ( N_{\tau} + 4).
\ee
The constant solutions obtained here are trivial ones to 
Heun's differential equations. However,
due to the rescaling factors in \eqref{rescale1} 
and \eqref{rescale2}, these correspond to
nontrivial eigenfunctions of the scalar Laplacian.
These are closely related to
holomorphic functions on the cone $C(X_5)$.
In the next section, we will discuss this point in detail.

\subsection{$n=1$}

Next, let us examine the first excited states ($n=1$).
The polynomial conditions \eqref{detM} for this case
\be
k_x ( k_x + Q_1^{(x)} ) - a_x \gamma_x P_1 = 0, \qq
k_y ( k_y + Q_1^{(y)} ) - a_y \gamma_y P_1 = 0,
\ee
yield the same algebraic equation for $\tilde{C}$:
\be
\tilde{C}^2 + 2 \nu \tilde{C}
+ \alpha \beta ( N_{\tau} + 3 ) 
( N_{\tau} - N_{\phi} - N_{\psi} + 1 ) = 0,
\ee
where
\be
\nu:= \frac{1}{2} \alpha ( N_{\tau} + 2 - N_{\phi})
+ \frac{1}{2} \beta( N_{\tau} + 2 - N_{\psi} ).
\ee
Therefore, if 
\be
\tilde{C} = - \nu
\pm \sqrt{\nu^2 - \alpha \beta ( N_{\tau} + 3)
( N_{\tau} - N_{\phi} - N_{\psi} + 1) },
\ee
then there exist polynomial solutions of degree one
for both $x$ and $y$
systems.
The eigenvalues of the scalar Laplacian are given by
\be
E = ( N_{\tau} + 2 )( N_{\tau} + 6 ).
\ee

\subsection{Comments for $n \geq 2$}

We comment on the cases for $n \geq 2$.
For $n=2$, we can show that
\be
(x_1 - x_2)^3 \det (M_x^{(2)} - k_x ) 
- (\alpha - \beta)^3 \det( M_y^{(2)} - k_y ) 
= 4 \mu ( N_{\tau} + 4 )( N_{\tau} + 5 ).
\ee
Therefore, $\det (M_x^{(2)} - k_x)$ and $\det (M_y^{(2)}-k_y)$ 
cannot vanish simultaneously, and so 
the polynomial solutions for both $x$ and $y$ systems
are not allowed in this case.

For $n=3$, the following relation holds
\be
\begin{split}
& ( x_1 - x_2 )^4 \det ( M_x^{(3)} - k_x ) 
- ( \alpha - \beta )^4 \det ( M_y^{(3)} - k_y ) \\
&= 12 \mu ( N_{\tau} + 6 )
\bigl( 2 ( N_{\tau} + 6) \tilde{C}
+ 3 ( \alpha + \beta) ( N_{\tau} + 4 ) ( N_{\tau} + 5)
- 3 ( N_{\tau} + 5 ) ( \alpha N_{\phi} + \beta N_{\psi} )
\bigr).
\end{split}
\ee
If there is a constant $\tilde{C}$ such that
both $\det (M_x^{(3)} - k_x)$ and $\det (M_y^{(3)} - k_y)$
vanish, 
the right-handed side of the equation above must be zero.
But it seems that
\be
\tilde{C} = - \frac{3 (N_{\tau} + 5)}{2(N_{\tau}+6)}
\bigl( ( \alpha + \beta) ( N_{\tau} + 4) 
- ( \alpha N_{\phi} + \beta N_{\psi} )
\bigr)
\ee
is not a solution of $\det(M_x^{(3)}-k_x) = 0$ or 
$\det(M_y^{(3)} - k_y)=0$. So there is no simultaneous polynomial
solution.

In general, for $n \geq 2$,
$\det(M_x^{(n)} - k_x)=0$ and $\det(M_y^{(n)}-k_y)=0$
give two different algebraic equations for one constant $\tilde{C}$. 
It seems that they do not have a common solution and 
hence there is no simultaneous polynomial solution of degree $n$
for both $x$ and $y$ systems.
We have to search for the Heun functions of non-polynomial type 
at least for one Heun's.

\section{Holomorphic functions and BPS mesons}

In subsection 4.2,
we determined the constant solutions to Heun's differential
equations. 
Up to a normalisation constant, 
the corresponding eigenfunctions 
are summarised as
\bel{grs}
\Psi^{(0)}[N] = \ex^{\im N^i \phi_i}\,  
( x - x_1 )^{\alpha_1} ( x_2 - x )^{\alpha_2}
( x_3 - x )^{\alpha_3} 
( 1 - y )^{\beta_1} ( 1 + y )^{\beta_2}
\bigl( 1 - (y/y_3) \bigr)^{\beta_3}.
\ee
In deriving the solutions above, we have assumed that
\bel{K1}
\alpha_1 \geq 0, \qq
\alpha_2 \geq 0, \qq
\beta_1 \geq 0, \qq
\beta_2 \geq 0.
\ee
With this assumption, the ground states \eqref{grs}
are regular in the regions
$x_1 \leq x \leq x_2$, $-1 \leq y \leq 1$.
Note that \eqref{K1} are sufficient conditions
for the eigenfunctions \eqref{grs} of the scalar Laplacian being regular.

Anyway, let us discuss the relation between the solution \eqref{grs}
and holomorphic functions.
Since $C(X_5)$ is a K\"{a}hler cone, 
any holomorphic function $w$ satisfies
the Laplace equation $\Box_{(6)} w = 0$. 
By restricting to the base space $X_5$,
the holomorphic function with the scaling dimension $\Delta$
becomes an eigenfunction of $\Box_{(5)}$ with
eigenvalue $E = \Delta ( \Delta + 4 )$.
The scaling dimension of the holomorphic function is defined by
$r(\partial/\partial r) w = \Delta w$.

In the symplectic approach to the Calabi-Yau cones, 
there is a way to construct the holomorphic coordinates 
by using a symplectic potential $G$. 
The holomorphic coordinates are given by \cite{MSY}
\be
w_i:= \mathrm{const} \times \exp\left(
\frac{\partial G(\xi)}{\partial \xi^i} + \im \phi_i \right),
\qq
i=1,2,3.
\ee
The coordinates $w_i$ are periodic functions
under the shift $\phi_i \rightarrow \phi_i + 2\pi$.
In these holomorphic coordinates, the holomorphic $(3,0)$-form
$\Omega$ is given by 
$\Omega = \de w_1 \wedge \de w_2 \wedge \de w_3/(w_2 w_3)$.

Since we know the symplectic potential \eqref{PotL}, 
we can write down the explicit form of the holomorphic coordinates
on $C(L^{a,b,c})$.
From \eqref{LG1}, and with an appropriate choice of the normalisation,
we have 
\be
\begin{split}
w_i &= r^{B_i} \ex^{\im \phi_i}
(x - x_1)^{(1/2) (v_1)_i} ( x_2 - x)^{(1/2)(v_3)_i}
( x_3 - x )^{(1/2) (v_5)_i } \\
& \qq \times
( 1 - y )^{(1/2) (v_2)_i } ( 1 + y )^{(1/2) (v_4)_i}
\left(1 - \frac{y}{y_3} \right)^{(1/2) (v_6)_i }.
\end{split}
\ee
The scaling dimension of $w_i$ is equal to the $i$-th component
of the Reeb vector $B$, and so
the dimension $\Delta_N$ of a holomorphic function 
$\Psi[N]:=w_1^{N^1} w_2^{N^2} w_3^{N^3}$
is given by
\be
\Delta_N = \langle B, N \rangle = B_i N^i = N_{\tau}.
\ee
Recalling \eqref{vdefe} and \eqref{vdefo},
we see that
the facets $\langle v_A, \xi \rangle = 0$ ($A=1,2,3,4$)
correspond to $x=x_1$, $y=1$, $x=x_2$ and $y=-1$, respectively. 
The necessary condition for the regularity of 
$\Psi[N]$ at the facets requires
\bel{nec}
\langle v_A, N \rangle \geq 0, \qq A=1,2,3,4.
\ee
Therefore, 
there exists a holomorphic function
to each integral lattice point $(N^1,N^2,N^3)$
in the polyhedral cone.
This fact is well-known in toric geometry.
Using \eqref{exx}, \eqref{exy} and \eqref{formN},
we can see that the restriction gives the eigenfunction
\eqref{grs},
\be
\Psi^{(0)}[N] = w_1^{N^1} w_2^{N^2} w_3^{N^3} \Bigr|_{r=1}.
\ee
The eigenvalue 
is given by $E= \Delta_N(\Delta_N +4)
= N_{\tau} ( N_{\tau} + 4)$, which is consistent with
\eqref{E0}.

The AdS/CFT implies that
these eigenfunctions correspond to the BPS mesonic operators
with the $R$-charge $(2/3) \Delta_N$.
For general $L^{a,b,c}$, it is difficult to treat all
integral points in the polyhedral cone.

One obvious integral point is
$(N^1, N^2, N^3) = (1,0,0) =: u_0.$
This vector lies in the polyhedral cone for any toric data.
Therefore, it is natural to identify its dual with
the short BPS meson operator $\mathcal{O}_{\beta}$
with the $R$-charge $2$, which exists for any toric
superconformal quiver gauge theory \cite{BH,BK05}.
The eigenfunction is given by
\be
\Psi^{(0)}[u_0]
= \ex^{\im \phi_1} \prod_{m=1}^3
(  x - x_m )^{1/2} \left( 1 - \frac{y}{y_m} \right)^{1/2}.
\ee

Other manifest possibilities are four vectors
pointing at the directions along 
the four edges of the polyhedral cone:
\be
\begin{split}
u_1 &:= v_1 \times v_2 = (b, -b, ak-1), \\
u_2 &:= v_2 \times v_3 =(a, b-c, -a(k+l) ), \\
u_3 &:= v_3 \times v_4 =(0, c, al ), \\
u_4 &:= v_4 \times v_1 =(0, 0, 1).
\end{split}
\ee
Note that they obey a linear relation:
\be
ac u_1 - bc u_2 + bd u_3 - ad u_4 = 0.
\ee
Here we assume that the vectors $u_i$ are primitive,
i.e., whose components do not have a common factor.
If these vectors are not primitive, which are
related to the orbifolds of the Sasaki-Einstein manifolds,
more careful treatment is needed. But 
we do not argue this subtle point.
The corresponding eigenfunctions are explicitly given by
\bel{EF4}
\begin{split}
\Psi^{(0)}[u_1]
&= \ex^{\im b \phi_1 - \im b \phi_2 + \im (ak-1) \phi_3} \\
& \qq \times 
(x_2-x)^{(1/2)d} ( x_3-x)^{(1/2) \langle v_5, u_1 \rangle} 
(1 + y )^{(1/2)b} 
\bigl( 1 -(y/y_3) \bigr)^{(1/2) \langle v_6, u_1 \rangle},\\
\Psi^{(0)}[u_2] 
&=\ex^{\im a \phi_1 + \im (b-c) \phi_2 - \im a(k+l) \phi_3} \\
& \qq \times
(x-x_1)^{(1/2)d} ( x_3 - x)^{(1/2) \langle v_5, u_2 \rangle} 
( 1 + y )^{(1/2)a}
\bigl( 1 - (y/y_3) \bigr)^{(1/2) \langle v_6, u_2 \rangle}, \\
\Psi^{(0)}[u_3] 
&=\ex^{\im c \phi_2 + \im al \phi_3} \\
& \qq \times
(x-x_1)^{(1/2)c} ( x_3 - x)^{(1/2) \langle v_5, u_3 \rangle} 
( 1 - y )^{(1/2)a}
\bigl( 1 - (y/y_3) \bigr)^{(1/2) \langle v_6, u_3 \rangle}, \\
\Psi^{(0)}[u_4]
&= \ex^{\im \phi_3} \\
& \qq \times 
(x_2-x)^{(1/2)c} ( x_3-x)^{(1/2) \langle v_5, u_4 \rangle} 
(1 - y )^{(1/2)b} 
\bigl( 1 -(y/y_3) \bigr)^{(1/2) \langle v_6, u_4 \rangle}.
\end{split}
\ee 

It is natural to identify primary operators related to
these four integral points $u_i$ $(i=1,2,3,4)$
with the four extremal BPS meson operators 
$\mathcal{O}_{LD}$, $\mathcal{O}_{RD}$, $\mathcal{O}_{RU}$
and $\mathcal{O}_{LU}$, respectively \cite{BK05}.
They also  
correspond to four $(p,q)$-branes \cite{BK05,FHMSVW}, 
whose $(p,q)$-charges can be read from the second 
and the third components of $u_i$ :
$(p,q) = (-(u_i)^2, -(u_i)^3)$.
Classical counterparts are
the four massless BPS geodesics considered in 
\cite{BK05}\footnote{The conditions $y=\pm 1$ in our notation seem
to correspond to the positions $y = \mp 1$ 
where the massless BPS geodesics stay \cite{BK05}.}.
Using the notation of \cite{FHMSVW} for the fundamental
chiral fields, the $R$-charges of the extremal BPS mesons are
related to the toric data as follows:
\[
R[\mathcal{O}_{LD} ] = d R[Z] + b R[U_2] = \frac{2}{3} 
\langle B, v_1 \times v_2 \rangle, \qq
R[ \mathcal{O}_{RD} ] = d R[Y] + a R[ U_2] = \frac{2}{3}
\langle B, v_2 \times v_3 \rangle,
\]
\bel{RC4}
R[\mathcal{O}_{RU}] = c R[Y] + a R[U_1] = \frac{2}{3}
\langle B, v_3 \times v_4 \rangle, \qq
R[\mathcal{O}_{LU} ] = c R[Z] + b R[ U_1] = \frac{2}{3}
\langle B, v_4 \times v_1 \rangle.
\ee
By comparing the powers of $(x-x_1)^{1/2}$, $(x_2-x)^{1/2}$,
$(1-y)^{1/2}$, $(1+y)^{1/2}$ in \eqref{EF4} 
with the coefficients of $R[Y]$, $R[Z]$, $R[U_1]$, 
$R[U_2]$ in \eqref{RC4}, we find the relation between the
Reeb vector and the $R$-charges of the distinguished 
bifundamental fields
$Y$, $U_1$, $Z$, $U_2$:
\be
\frac{2}{3} B = R[Y] v_1 + R[U_1] v_2 + R[Z] v_3 + R[U_2] v_4.
\ee
The Reeb vector $B$ can be determined by the $Z$-minimisation,
while the $R$-charges can be computed by the $a$-maximisation \cite{IW}.
This relation connects the charges in the gauge theory to
the geometric object $B$, and is a consequence of the AdS/CFT correspondence.

\section{Summary and Discussion}

In this paper, we have presented
the symplectic potentials 
for a wide class of toric Sasaki-Einstein manifolds.

The spectrum of the scalar Laplacian for 
$L^{a,b,c}$ is investigated.
The eigenvalue problem 
leads to two Heun's differential equations, which are
correlated through one constant $\tilde{C}$.
We find that the exponents at the regular singularities
can be nicely characterised by the toric data.

What is the physical interpretation of $\tilde{C}$? 
The eigenvalue equations 
are separated
into five ordinary differential equations.
There are four manifestly constant quantities. Three 
are related to the three $U(1)$ isometries.
The fourth is $E$, which is the ``energy.''
The physical meaning of these four constants are clear.
The fifth constant $\tilde{C}$ appears 
through the separation of variables for $x$ and $y$.
$\tilde{C}$ is not a Noether charge.
Originally, the $L^{a,b,c}$ metrics were reduced from 
the Kerr-AdS black holes \cite{CLPP,CLPP2}.
This way of generating Sasaki-Einstein manifolds was
started out with \cite{HSY}, where the special case $Y^{p,q}$ was
studied. 
The integrability property of geodesics and eigenvalue equations
on the black holes is 
guaranteed by the existence of a symmetric rank two
Killing tensor \cite{KL}. Therefore, there may be 
a connection between the constant $\tilde{C}$
and the Killing tensor.

The ground states
can be obtained as a restriction of the holomorphic
functions on the Calabi-Yau cone.
Some families 
of holomorphic functions were constructed for $Y^{p,q}$\cite{BHOP}.
But the analysis was done in rather ad hoc manner.
By combining knowledge of the explicit symplectic potential
\eqref{PotL}
and the relationship between the exponents and the toric data \eqref{exx}, 
\eqref{exy}, we have shown that the eigenfunctions \eqref{grs}
or the holomorphic functions 
$w_1^{N^1} w_2^{N^2} w_3^{N^3}$ have one-to-one correspondence
with integral points $(N^1,N^2,N^3)$ in the convex polyhedral cone.
The holomorphic functions have the scaling dimensions $B_i N^i$.
The corresponding values of $R$-charges $(2/3) B_i N^i$
are consistent with the results of the dual quiver gauge theories.
For certain BPS operators, the corresponding integers
$(-N^2, -N^3)$ are found to be equal
to the $(p,q)$-charges of the $(p,q)$-web 
of $5$-branes \cite{PQW}.
Moreover, the powers of $(x-x_1)^{1/2}$, $(x_2-x)^{1/2}$,
$(1-y)^{1/2}$, $(1+y)^{1/2}$ are closely related to
the numbers of the
``constituent'' bifundamental fields $Y$, $Z$, $U_1$, $U_2$,
respectively. 
So there may be more physical explanation of these
eigenfunctions.

In addition to the ground states, we constructed
the first excited eigenfunctions.
Due to the difficulties for obtaining the Heun functions,
the determination of the full spectrum for $L^{a,b,c}$ 
is still an open problem.

\vspace{10mm}
\noindent
{\bf \Large Acknowledgements}

\vspace{5mm}

We would like to thank Y. Hashimoto, H. Kihara and M. Sakaguchi
for discussions.
This work is supported by the 21 COE program 
``Constitution of wide-angle mathematical basis focused on knots.''
The work of Y.Y. is supported by 
the Grant-in Aid for Scientific Research
(No. 17540262 and No. 17540091) from Japan Ministry of Education.

\appendix

\section{Details on the symplectic potential for $L^{a,b,c}$}

In this appendix, 
we show that the symplectic potential \eqref{PotL}
is indeed a solution to the Monge-Amp\`{e}re equation
for $L^{a,b,c}$.
The key relations, obtained from \eqref{vdefe}, \eqref{vdefe2} 
and \eqref{vdefo}, are 
\bel{vid}
\frac{ \langle v_1, \xi \rangle}{x-x_1}
+ \frac{\langle v_3, \xi \rangle}{x-x_2}
+ \frac{\langle v_5, \xi \rangle}{x-x_3}=0,
\qq
\frac{\langle v_2, \xi \rangle}{y-1}
+ \frac{\langle v_4, \xi \rangle}{y+1}
+ \frac{\langle v_6, \xi \rangle}{y-y_3} = 0.
\ee
Using these relations, we have (up to an irrelevant constant)
\bel{LG1}
\begin{split}
\frac{\partial G}{\partial \xi^i}
&= \frac{1}{2} B_i \log \langle B, \xi \rangle \\
& + \frac{1}{2} (v_1)_i \log | x-x_1|
+ \frac{1}{2} (v_3)_i \log | x-x_2|
+ \frac{1}{2} (v_5)_i \log|x-x_3| \\
& + \frac{1}{2} (v_2)_i \log | y - 1|
+ \frac{1}{2} (v_4)_i \log|y+1|
+ \frac{1}{2} (v_6)_i \log| y - y_3|.
\end{split}
\ee
Recall that $B_1 = 3$, and 
$(v_I)_1 = 1$, for $I= 1,2,\dotsc, 6$.
Then \eqref{LG1} for $i=1$ leads to
\bel{Gd1}
2 \frac{\partial G}{\partial \xi^1}
= \log \left| \langle B, \xi \rangle^3 
(x-x_1)(x-x_2)(x-x_3)( 1-y^2)(y-y_3) \right|.
\ee
By differentiating \eqref{LG1} once more, we have
\be
\begin{split}
\frac{\partial^2 G}{\partial \xi^i \partial \xi^j} 
&= \frac{B_i B_j}{2 \langle B, \xi \rangle} 
+ \frac{1}{2}
\left[ \frac{(v_1)_i}{x-x_1}
+ \frac{(v_3)_i}{x-x_2} + \frac{(v_5)_i}{x-x_3} \right]
\frac{\partial x}{\partial \xi^j} \\
& \qq \qq + \frac{1}{2}
\left[ \frac{(v_2)_i}{y-1}
+ \frac{(v_4)_i}{y+1}
+ \frac{(v_6)_i}{y-y_3} \right] \frac{\partial y}{\partial \xi^j}.
\end{split}
\ee
Contraction with $\de \xi^i \de \xi^j$ yields
\bel{Gdd}
\begin{split}
\frac{\partial^2 G}{\partial \xi^i \partial \xi^j}
 \de \xi^i  \de \xi^j 
&= \frac{(\de \langle B, \xi \rangle)^2}{2 \langle B, \xi \rangle} 
+ \frac{1}{2} \left[
\frac{\de \langle v_1, \xi \rangle}{x-x_1} 
+ \frac{\de \langle v_3, \xi \rangle}{x-x_2}
+ \frac{\de \langle v_5, \xi \rangle}{x-x_3} \right] \de x \\
& \qq \qq \qq + \frac{1}{2} \left[
\frac{\de \langle v_2, \xi \rangle}{y-1}
+ \frac{\de \langle v_4, \xi \rangle}{y+1}
+ \frac{\de \langle v_6, \xi \rangle}{y-y_3} \right] \de y.
\end{split}
\ee
From \eqref{vid}, the following relations can be obtained
\bel{iden1}
\frac{\de \langle v_1, \xi \rangle}{x-x_1}
+ \frac{\de \langle v_3, \xi \rangle}{x-x_2}
+ \frac{\de \langle v_5, \xi \rangle}{x-x_3}
=\left[ \frac{\langle v_1, \xi \rangle}{(x-x_1)^2}
+ \frac{\langle v_3, \xi \rangle}{(x-x_2)^2}
+ \frac{\langle v_5, \xi \rangle}{(x-x_3)^2} \right] \de x
= \frac{r^2 \rho^2}{2 \Delta_x} \de x,
\ee
\bel{iden2}
\frac{\de \langle v_2, \xi \rangle}{y-1} 
+ \frac{\de \langle v_4, \xi \rangle}{y+1}
+ \frac{\de \langle v_6, \xi \rangle}{y-y_3}
= \left[ \frac{ \langle v_2, \xi \rangle}{(y-1)^2}
+ \frac{\langle v_4, \xi \rangle}{(y+1)^2}
+ \frac{\langle v_6, \xi \rangle}{(y-y_3)^2} \right]
\de y
= \frac{r^2 \rho^2}{2\Delta_y}
\de y.
\ee
Here
\be
\Delta_x = (x-x_1)(x-x_2)(x-x_3), \qq
\Delta_y = \frac{1}{2} (1-y^2)\bigl( (\alpha + \beta) 
+ (\alpha - \beta) y \bigr).
\ee
Substitution of \eqref{iden1} and \eqref{iden2}
into \eqref{Gdd} yields
\bel{GtoL}
\frac{\partial^2 G}{\partial \xi^i
\partial \xi^j} \de \xi^i \de \xi^j 
= \de r^2 + \frac{r^2 \rho^2}{4\Delta_x} \de x^2
+ \frac{r^2 \rho^2}{4\Delta_y} \de y^2. 
\ee
Note that the Jacobian of the coordinate transformation 
\eqref{vdefe} from $(\xi^1,\xi^2,\xi^3)$ to $(r,x,y)$ 
is proportional to $r^5 \rho^2$.
From \eqref{GtoL} and the Jacobian, one can show that
\be
\det\left( \frac{\partial^2 G}{\partial \xi^i \partial \xi^j}
\right) \propto \frac{1}{r^6 \Delta_x \Delta_y}
\propto \frac{1}{\langle B, \xi \rangle^3
(x-x_1)(x-x_2)(x-x_3)(1-y^2)(y-y_3)}.
\ee
Hence, with \eqref{Gd1}, we have
\be
\det\left( \frac{\partial^2 G}{\partial \xi^i \partial \xi^j}
\right) = \mathrm{const} \times
\exp\left( - 2 \frac{\partial G}{\partial \xi^1} \right),
\ee
which implies that \eqref{PotL} is a solution to the Monge-Amp\`{e}re
equation.
Moreover, \eqref{GtoL} implies that the symplectic potential
\eqref{PotL} indeed reproduces the $L^{a,b,c}$ metric \eqref{Lmet}
by setting $y=\cos 2\theta$.


\end{document}